\documentclass[%
 reprint,
 superscriptaddress,
 amsmath,amssymb,
 aps,
]{revtex4-1}

\usepackage[dvipdfmx]{graphicx}% Include figure files
\usepackage{dcolumn}% Align table columns on decimal point
\usepackage{bm}% bold math
\usepackage{multirow}

\begin{document}

\preprint{APS/123-QED}

\title{EUV and Visible Spectroscopy of Promethiumlike Heavy Ions}% Force line breaks with \\
%\thanks{A footnote to the article title}%

\author{Yusuke Kobayashi}
\affiliation{Institute for Laser Science, The University of Electro-Communications, Chofu, Tokyo 182-8585, JAPAN}

\author{Kai Kubota}
\affiliation{Institute for Laser Science, The University of Electro-Communications, Chofu, Tokyo 182-8585, JAPAN}

\author{Kazuki Omote}
\affiliation{Institute for Laser Science, The University of Electro-Communications, Chofu, Tokyo 182-8585, JAPAN}

%\author{Takayuki Nakajima}
%\affiliation{Institute for Laser Science, The University of Electro-Communications, Chofu, Tokyo 182-8585, JAPAN}

\author{Akihiro Komatsu}
\affiliation{Institute for Laser Science, The University of Electro-Communications, Chofu, Tokyo 182-8585, JAPAN}

\author{Junpei Sakoda}
\affiliation{Institute for Laser Science, The University of Electro-Communications, Chofu, Tokyo 182-8585, JAPAN}

\author{Maki Minoshima}
\affiliation{Institute for Laser Science, The University of Electro-Communications, Chofu, Tokyo 182-8585, JAPAN}

\author{Daiji Kato}
\affiliation{National Institute for Fusion Science, Toki, Gifu 509-5292, JAPAN}
\affiliation{Department of Fusion Science, SOKENDAI (The Graduate University of Advanced Studies), Toki, Gifu 509-5292, JAPAN}

\author{Jiguang Li}
\affiliation{Institute of Applied Physics and Computational Mathematics, Beijing 100088, CHINA}

\author{Hiroyuki A. Sakaue}
\affiliation{National Institute for Fusion Science, Toki, Gifu 509-5292, JAPAN}

\author{Izumi Murakami}
\affiliation{National Institute for Fusion Science, Toki, Gifu 509-5292, JAPAN}
\affiliation{Department of Fusion Science, SOKENDAI (The Graduate University of Advanced Studies), Toki, Gifu 509-5292, JAPAN}

\author{Nobuyuki Nakamura}
% \email{n\_nakamu@ils.uec.ac.jp}
\affiliation{Institute for Laser Science, The University of Electro-Communications, Chofu, Tokyo 182-8585, JAPAN}

\date{\today}% It is always \today, today,
             %  but any date may be explicitly specified

\begin{abstract}
We present extreme ultraviolet and visible spectra of promethiumlike tungsten and gold obtained with an electron beam ion trap (EBIT).
Although the contributions from a few charge states are involved in the spectra, the charge state of the ion assigned to the observed lines is definitely identified by the time-of-flight analysis of the ions performed at the same time with the spectroscopic measurements.
Experimental results are compared with collisional-radiative model calculations as well as previous experimental and theoretical studies.
%\begin{description}
%\item[Usage]
%Secondary publications and information retrieval purposes.
%\item[PACS numbers]
%\pacs{32.30.-r, 31.15.ag, 52.20.-j}

%May be entered using the \verb+\pacs{#1}+ command.
%\item[Structure]
%You may use the \texttt{description} environment to structure your abstract;
%use the optional argument of the \verb+\item+ command to give the category of each item. 
%\end{description}
\end{abstract}

%\pacs{32.30.-r, 31.15.ag, 52.20.-j}% PACS, the Physics and Astronomy
                             % Classification Scheme.
%\keywords{Suggested keywords}%Use showkeys class option if keyword
                              %display desired
\maketitle

%\tableofcontents

\section{\label{sec:introduction}Introduction}
The promethiumlike sequence (61 electron system) is one of the most important isoelectronic sequences because it has an alkali-metal structure with a closed $4f$ shell and one valence $5s$ electron for heavy elements.
The investigation of promethiumlike ions was triggered by the theoretical study by Curtis and Ellis in 1980, which first predicted the alkali-metal structure and its $5s$ -- $5p$ resonance lines~\cite{Curtis1,Ellis1}.
After the prediction, experimentalists have tried to find the resonance lines in the extreme ultraviolet (EUV) region of beam-foil~\cite{Johnson5,Theodosiou1,Trabert4,Kaufman2}, tokamak~\cite{Fournier1}, charge-exchange~\cite{Andersson1}, and electron beam ion trap (EBIT)~\cite{Hutton2,Wu1} spectra.
However, all of them failed to observe the resonance line definitely.
Recently, we clarified the reason for the lost resonance line through the observation of promethiumlike bismuth (the atomic number $Z=83$) with a compact EBIT and a collisional-radiative (CR) model analysis~\cite{Kobayashi1}.
In that study, we showed that the population of the $4f^{13}5s^2$ metastable state dominates over that of the $4f^{14}5s$ ground state in a plasma with a practical electron density, and as a result, the resonance lines are much less prominent compared to the $4f^{13}5s^2$ -- $4f^{13}5s5p$ transitions.
Our observation has recently been confirmed by the subsequent study with the Heidelberg EBIT~\cite{Bekker1}.
They observed the $4f^{14}5s$ -- $4f^{14}5p_{3/2}$ line in promethiumlike rhenium, osmium, iridium, and platinum ($Z$=75 to 78), but its intensity was confirmed to be much smaller than that of the $4f^{13}5s^2$ -- $4f^{13}5s5p$ transitions even for promethiumlike platinum, whose ground state is $4f^{14}5s$.

In this paper, we present spectra of promethiumlike tungsten ($Z=74$) and gold (79) not only to extend the previous studies for other atomic numbers but also to provide important atomic data for applications.
Gold is frequently used as a material of a cavity called hohlraum in inertial confinement fusion experiments~\cite{Lindl1}.
A hohlraum plays a key role to convert the focused laser beams into x-ray photons that heat the fusion target.
Thus the emission property of highly charged gold ions in the soft to hard x-ray regions is important to achieve efficient ignition.
On the other hand, tungsten is important material for magnetic confinement fusion reactors.
Since tungsten has a melting point that is the highest of any of metals and a low tritium retention, it is a potential candidate for the plasma facing material of fusion reactors, such as International Thermonuclear Experimental Reactor (ITER)~\cite{Pitts1}.
One of the key issues for successful operation of ITER is to diagnose and control the influx of tungsten sputtered from the wall into the core plasma.
Spectroscopic data of highly charged tungsten ions are thus strongly required to perform the spectroscopic diagnostics of the ITER plasma accurately.

Promethiumlike gold and tungsten have also been studied so far by beam-foil spectroscopy~\cite{Trabert4,Theodosiou1,Kaufman2,Johnson5} and EBIT spectroscopy~\cite{Hutton2,Wu1,Zhao1}.
However, in the beam-foil spectroscopy, rather complex spectra containing contributions from several charge states prevent detailed analysis and identification of the observed spectra.
In contrast, an EBIT is a useful device to obtain simple spectra containing contributions from only a few charge states.
In fact, several lines of promethiumlike heavy ions have been identified through the spectroscopic observation with an EBIT~\cite{Hutton2,Kobayashi1,Zhao1,Bekker1}.
In their studies, the charge state of the ion assigned to the observed lines was identified through the electron energy dependence of line intensity, which generally shows a gradual increase  just after the electron energy exceeds the ionization energy of the previous charge state and a gradual decrease when the electron energy is further increased beyond the ionization energy of the corresponding ion.
Such an identification has been proved to be reliable to some extent~\cite{Komatsu1,Nakamura19,Yatsurugi1}, but the assigned charge states may have an uncertainty due to some reasons: (1) the charge exchange contribution is difficult to estimate, (2) ionization energy is not accurately known, and (3) the electron energy can have a width comparable to the ionization energy separation.
The uncertainty is serious especially for relatively low charge state ions because the ionization energy is difficult to estimate due to its complex multi-electron structure, and its ionization-energy interval between adjacent charge states can be comparable to the uncertainty in the ionization energy and the electron energy.
On the other hand in the present study, the time-of-flight (TOF) analysis is used for definite identification of the ionic charge.
In addition to the EUV range, we also show spectra in the visible range, which contains magnetic dipole (M1) transitions between fine structure levels of the ground and metastable states.

\section{\label{sec:experiment}Experiment}
%\subsection{\label{subsec:CoBIT} compact electron beam ion trap (CoBIT)}
In the present study, highly charged tungsten and gold ions were produced with a compact electron beam ion trap (called CoBIT) at The University of Electro-Communications in Tokyo.
The detailed description on the device is given in the previous paper~\cite{cobit}.
Briefly, it consists essentially of an electron gun, a drift tube, an electron collector, and a high-critical-temperature superconducting magnet.
The drift tube is composed of three successive cylindrical electrodes (DT1, 2 and 3) that act as an ion trap by applying a positive potential (typically 30 V) at both ends (DT1 and 3) with respect to the middle electrode (DT2).
It is noted that the DT2 electrode of CoBIT was originally composed of six poles~\cite{cobit}, but the present DT2 is a cylindrical electrode with six slits for spectroscopic observation.
The electron beam emitted from the electron gun is accelerated towards the drift tube while it is compressed by the axial magnetic field (typically $\sim 0.08$ T) produced by the magnet surrounding the drift tube.
The compressed high-density electron beam ionizes the ions trapped in the drift tube.
Tungsten was introduced into the trap through a gas injector as a vapor of W(CO)$_6$, whereas gold was introduced as a vapor from an effusion cell~\cite{Yamada1} operated at 1100~$^\circ$C.

%\subsection{\label{subsec:spectroscopy} spectroscopic observations}
Emission from the trapped ions in the EUV range was observed with a grazing-incidence flat-field spectrometer~\cite{Ohashi4}.
The spectrometer consisted of an aberration-corrected concave grating with a groove number of 1200~mm$^{-1}$ (Hitachi 001-0660) and a Peltier-cooled back-illuminated CCD (Roper PIXIS-XO: 400B).
For the experiments reported herein, no entrance slit was used because CoBIT constitutes a line source that serves the same function as that of a slit.
The spectral resolution of this arrangement was typically 0.03~nm, which was mainly limited by the electron-beam width.
The wavelength was calibrated by using well-known transitions in Fe IX -- XIV~\cite{NISTdatabase2015} measured separately by injecting a Fe(C$_5$H$_5$)$_2$ vapor into CoBIT.
The uncertainty in the wavelength calibration was estimated to be 0.01~nm.

Emission in the visible range was observed with a commercial Czerny-Turner spectrometer.
A biconvex lens was used to focus the emission on the entrance slit of the spectrometer.
The diffracted light was detected by a back-illuminated CCD (Andor iDus 416) operated at -70~$^\circ$C.
The wavelength was calibrated using emission lines from several standard lamps placed outside \mbox{CoBIT}.
The uncertainty of the wavelength calibration was estimated to be 0.05~nm including systematics.

%\subsection{\label{subsec:TOF}time of flight measurements}

%In our previous spectroscopic studies with CoBIT, the charge state assigned to previously unreported lines has been identified based on the appearance energy through the observation of the electron energy dependence.
%For example, when a line is observed with an electron energy above the ionization energy of A$^{(q-1)+}$ while it is not observed with an energy below the ionization energy, the attribution of the line is assigned to A$^{q+}$.
% as already described in Sec.~\ref{sec:introduction}.
%The main reasons for the uncertainty are (1) poor-accuracy in the calculated ionization energy, (2) small ionization cross section near the threshold energy, and (3) electron energy width.

For the TOF analysis of the ionic charge, we have installed an ion extraction line in the present study.
It consists of an einzel lens, an electrostatic 90$^\circ$ bender, and a micro channel plate (MCP) detector.
The lens was installed in the bore of the liquid nitrogen server in the CoBIT chamber, whereas the bender and the MCP were installed just above the chamber.
The lens was used to focus the extracted ions into the entrance of the bender, and the bender was to separate the extracted ions from other particles, such as electrons and photons.
The signal from the MCP anode was directly observed with a digital oscilloscope without using any pre-amplifier.
The detailed description on the ion extraction system will be given elsewhere.

In the present TOF measurements, the drift tube was biased at +400 V, which served as an acceleration voltage for extracted ions (the operation mode with a biased drift tube is refereed to as ``ion extraction mode" hereafter).
It is in contrast with our previous spectroscopic studies where the drift tube potential has been fixed at ground (the operation mode with a grounded drift tube is refereed to as ``spectroscopy mode" hereafter).
Ion extraction was done by applying a ramp potential to DT2.
The hight and width of the ramp was 500 V and 10 $\mu$s, respectively.

%\section{\label{sec:crmodel}Collisional-Radiative Model Calculation}
\section{\label{sec:theory}Theoretical Calculations}

In order to analyze the EUV spectra of the promethiumlike tungsten, gold, and bismuth ions, we performed CR-model calculations with a set of the electron configurations $4f^{14}nl$, $4f^{13}5snl$, $4f^{13}5p^2$, $4f^{13}5p5d$, $4f^{12}5s^2 5l$, and $4d^9 4f^{14}5s^2$ ($n=5$ and 6, $l=0$ -- 4) consisting of 980 fine structure levels.
Populations of the excited levels were obtained by solving quasi-stationary-state rate equations with the CR model including electron collisions (excitation, de-excitation and ionization) and radiative decays (electric-dipole, -quadruple and -octupole, and magnetic-dipole and -quadruple).
Because the electron-beam energy was below the first-ionization energy of each ion, recombination from higher charge states was omitted.
Energy levels, electron collision strengths, and radiative decay rates were calculated by using the HULLAC code (version 7)~\cite{Shalom1} based on fully relativistic wave functions, and the collision strengths are calculated in the distorted wave approximation.
In order to improve the wavelengths of the $5s$--$5p$ resonance transitions in our previous work~\cite{Kobayashi1}, orbital wave functions optimized to different parametric potentials were used in the present calculations.

\if0
We analyzed the spectra of the promethiumlike ions of W$^{13+}$, Au$^{18+}$, and Bi$^{22+}$ by CR-model calculations with a set of excited levels, consisting of 980 fine-structure levels of $4f^{14}nl$, $4f^{13}5snl$, $4f^{13}5p^2$, $4f^{13}5p5d$, $4f^{12}5s^2 5l$, and $4d^9 4f^{14}5s^2$ for promethiumlike ions ($n=5$ and 6, $l=0$--4).
With the CR model, excited-level populations were obtained by solving quasi-stationary-state rate equations that account for electron collisions (excitation, de-excitation and ionization) and radiative decays (electric-dipole, -quadruple and -octupole, and magnetic-dipole and -quadruple).
Recombination from higher charge states was omitted because the electron-beam energy was below the first-ionization energy of each ion.
Atomic data (i.e. energy levels, electron collision strengths, and radiative decay rates) were calculated by using the HULLAC code (version 7)~\cite{Shalom1}.
The HULLAC calculations are based on fully relativistic wave functions, and the collision strengths are calculated in the distorted-wave approximation.
In order to improve the wavelengths of the$5s$--$5p$ resonance transitions in our previous work~\cite{Kobayashi1}, orbital wave functions optimized to different parametric potentials were used in the present calculations.
\fi

\begin{figure}[t]
\includegraphics[width=0.48\textwidth]{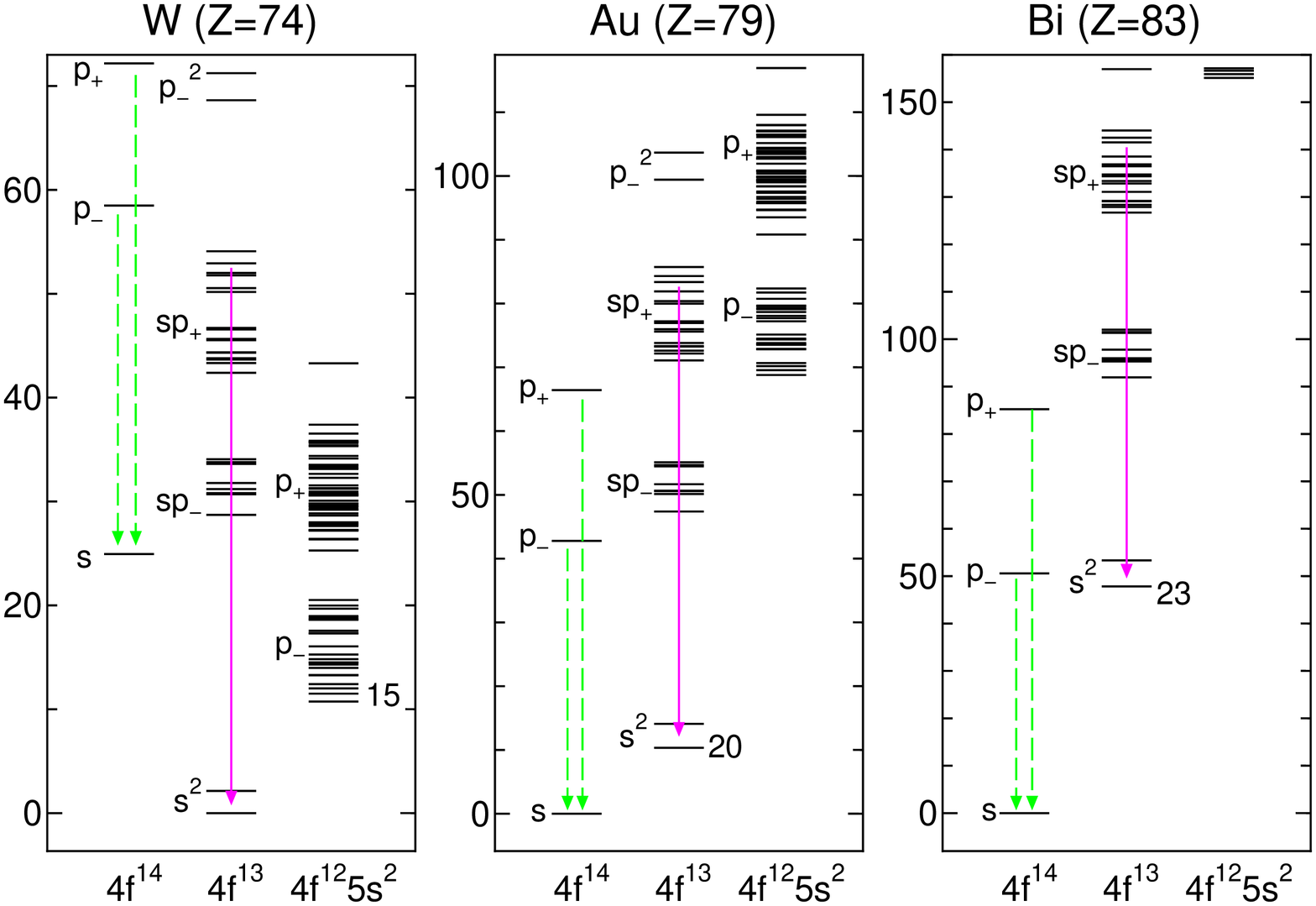}
\caption{\label{fig:levels_pm}
Calculated energy levels of promethiumlike tungsten, gold, and bismuth ions in the energy range below $E_{\rm ion}/4$, where $E_{\rm ion}$ is the ionization energy.
The principal quantum number $n=5$ is omitted, e.g. ``$sp\_$" represents $5s5p_{1/2}$.
The $5s^2$ -- $5s5p_{3/2}$ transitions, which are prominent in the region of current interest, are shown by the solid arrows, whereas the alkali-metal-like $5s$ -- $5p$ transitions are shown by the dashed arrows.
For the long-lived metastable states, the relative abundance with respect to the ground state calculated for an electron density of $10^{10}$ cm$^{-3}$ is indicated besides the level (total abundance for the lowest four levels are indicated for tungsten).
}
\end{figure}

The calculated energy levels for promethiumlike tungsten, gold, and bismuth are shown in Fig.~\ref{fig:levels_pm}.
%\begin{figure}[t]
%\includegraphics[width=0.45\textwidth]{levels_sm}
%\caption{\label{fig:levels_sm}
%Calculated energy levels of samariumlike ions ($n_e=62$).
%}
%\end{figure}
According to the present calculations,  the ground state configuration of promethiumlike gold and bismuth is $4f^{14}5s$, and the first excited level is [$4f^{13}5s^2$]$_{J=7/2}$, which can only decay via an electric octupole (E3) transition with a long decay lifetime (14~days for gold and 58~s for bismuth).
Thus, as already shown in our previous study~\cite{Kobayashi1}, the population in plasma is likely to be trapped in this metastable state over a wide electron density range.
In deed, the present calculations show the population of the metastable state is an order of magnitude larger than that of the ground state for an electron density of $10^{10}$ cm$^{-3}$.
When the atomic number decreases from 79 (gold) to 74 (tungsten), the collapse of the $4f$ orbital relaxes, so that $4f^{13}5s^2$ takes over from $4f^{14}5s$ as the ground state.
In the meantime, the $4f^{12}5s^2 5p_{1/2}$ configuration, which has E1 decay channels to $4f^{13}5s 5p_{1/2}$ for gold and bismuth, becomes the lowest excited configuration for tungsten.
Thus, the [$4f^{12}5s^2 5p_{1/2}$]$_J$ level in tungsten loses any E1 decay channel, and as a result acts as a metastable state especially when it has a large $J$.
For example the lowest level [$4f^{12}5s^2 5p_{1/2}$]$_{J=11/2}$ has a decay lifetime of 1.6~s.
Thus, the population in promethiumlike tungsten tends to be trapped in the [$4f^{12}5s^2 5p_{1/2}$]$_J$ levels.
The present calculation shows that the population of the [$4f^{12}5s^2 5p_{1/2}$]$_J$ levels is larger than that of the ground state for the lowest four levels with $J=7/2$ to 11/2.

The transition lifetime and corresponding population can be affected by the magnetic field and the hyperfine interaction, which are not considered in the present CR model calculation.
For example for promethiumlike gold and bismuth, the [$4f^{13}5s^2$]$_{J=7/2}$ metastable state is mixed with [$4f^{13}5s^2$]$_{J=5/2}$ in the presence of an external magnetic field, and as a result, the decay lifetime of the metastable state can be modified.
However, the [$4f^{13}5s^2$]$_{J=5/2}$ state decays to the ground state only via E3, whose transition probability is similar to that of the [$4f^{13}5s^2$]$_{J=7/2}$ metastable state.
Thus, the effect of the magnetic field on the decay lifetime of the metastable state is considered to be negligible.
On the other hand, hyperfine interaction can mix the [$4f^{13}5s^2$]$_{J=7/2}$ metastable state not only with [$4f^{13}5s^2$]$_{J=5/2}$ via the magnetic dipole hyperfine interaction but also with [$4f^{14}5p$]$_{J=3/2}$ via the electric quadrupole hyperfine interaction.
Since the [$4f^{14}5p$]$_{J=3/2}$ state can decay to the ground state via an E1 transition, the mixing would quench the population of the metastable state.
We have estimated this effect for the [$4f^{13}5s^2$]$_{J=7/2}$ metastable state of promethiumlike bismuth, which has a nuclear spin of 9/2.
The calculated result for the hyperfine induced transition rate is $10^{-7}$~s$^{-1}$, which is considerably smaller than the E3 transition rate of the [$4f^{13}5s^2$]$_{J=7/2}$ state.
Consequently, we consider that the effects of the magnetic field and hyperfine interaction are negligible for the system studied in this paper.
The details of the theoretical method and result will be published elsewhere.

For the fine structure splitting between $4f^{13}5s^2$ $J=7/2$--5/2 and the M1 transition rate, elaborate calculations using GRASP2K (A General-Purpose Relativistic Atomic Structure Program)~\cite{Jonsson2} were performed.
The calculations are based on the multi-configuration Dirac-Hartree-Fock (MCDHF) method and the relativistic configuration interaction (RCI) method.
In the MCDHF method, wave functions of the fine-structure states, $\Psi$, are expressed as a linear combination of anti-symmetrized configuration state functions (CSFs), $\Phi$,
\begin{equation}
\label{eq:MCDHF}
\Psi(\Gamma P J M)=\sum_r c_{r\Gamma} \Phi(\gamma_r P J M) .
\end{equation}
In the expression above, $J$ and $M$ are the angular quantum numbers, $P$ denotes parity, and $\gamma_r$ represents all information to specify the configuration state such as orbital occupation numbers, coupling, seniority numbers, etc.
The CSFs are coupled products of one-electron Dirac orbitals.
Optimization of each orbital wave function was performed in the extended optimal level (EOL) scheme.
In the present RCI calculations, the Breit interaction, vacuum polarization, and self-energy corrections are taken into account as perturbation.

\section{\label{sec:results}Results and discussion}

\begin{figure}[t]
\includegraphics[width=0.4\textwidth]{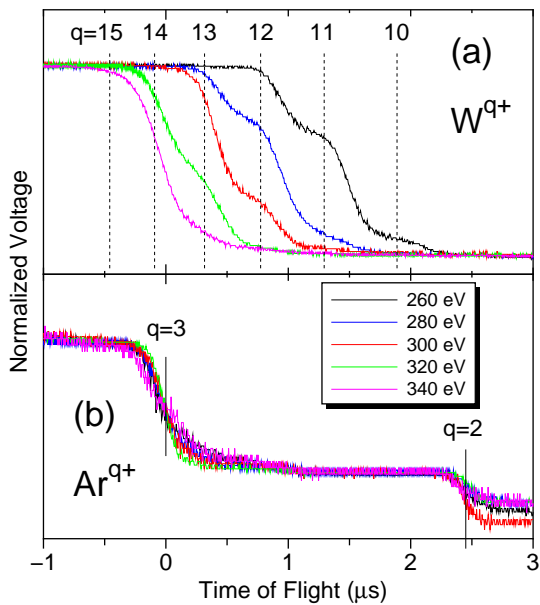}
\caption{\label{fig:TOF_W}
Output signals from the MCP anode for (a) tungsten and (b) gold ions extracted from the compact EBIT.
The horizontal axis is the time with respect to the flight time of Ar$^{3+}$.
The vertical axis is the voltage observed with a digital oscilloscope directly connected to the MCP.
The solid vertical lines in (b) correspond to the position (peak position of the derivative) of Ar$^{3+}$ and Ar$^{2+}$ ions, whereas the dotted lines in (a) indicate the flight time of tungsten ions estimated from that of the Ar ions.
}
\end{figure}

Figure~\ref{fig:TOF_W}(a) shows typical MCP output signals of extracted tungsten ions observed at electron energies of 260 -- 340~eV.
Here the electron energy just represents the potential difference between the cathode and the DT2 electrode (middle of the ion trap).
The vertical axis represents the voltage observed with an oscilloscope directly connected to the MCP anode.
Arrival of an extracted ion bunch is confirmed as a falling edge in the figure.
It is noted that the voltage of each signal was normalized to have the same level before and after the ion detection.
The time resolution (the falling time of the signal) is mainly limited by the broadening of the ion bunch.
As seen in the figure, a couple of falling edges were observed for each electron energy, which reflects the fact that only a couple of charge states were involved for each energy.
For identifying the charge state for each falling edge, argon was injected into CoBIT as a reference of the horizontal axis corresponding to (the square root of) the mass-to-charge ratio.
Figure~\ref{fig:TOF_W}(b) shows the signals obtained for the extracted argon ions.
For each electron energy, both argon and tungsten were extracted with completely the same operational parameters except for the ion extraction frequency.
For tungsten ions, the TOF measurements were performed at the same time with the spectroscopic measurements described later, and the trapped ions were dumped 
every 30~s.
On the other hand, argon ions were dumped with a frequency in the order of 10 Hz to produce relatively low charged ions such as Ar$^{2+}$ effectively.
The ions extracted from the drift tube traveled through the electron collector, whose potential was changed depending on the electron energy because it was biased at a fixed potential with respect to the cathode.
Thus the ion trajectory could differ depending on electron energy, which was controlled by the cathode potential.
Actually, it was observed that the arrival time of extracted argon ions slightly shifted depending on the electron energy even though completely the same operational parameters were used other than the cathode potential.
Thus the signal for each electron energy shown in Fig.~\ref{fig:TOF_W} was shifted so that the falling edge in the argon signal coincides with each other.
%The falling edge positions for tungsten ions estimated from the position of the argon ions are shown by the dotted lines in Fig.~\ref{fig:TOF_W}(a).
The dotted lines in Fig.~\ref{fig:TOF_W}(a) indicate the TOF of tungsten ions estimated from that of argon ions assuming a tungsten mass of 183.8 (average mass weighted by the natural abundance).
It is noted that the TOF differences among different isotopes are enough small to separate the charge state.
As seen in the figure, slight shift is found between the estimated position and the observed falling edge.
It probably reflects the fact that highly charged heavy ions were extracted from a potential deeper than that for low charged light ions, so that the arrival time of highly charged heavy ions was slightly delayed compared with that of low charged light ions even if they have the same mass-to-charge ratio.
Considering this slight delay, it is confirmed that the charge state of the ions was dominated with 13+ (promethiumlike) when the electron energy was 300~eV although some fraction of 12+ (samariumlike) is confirmed.
Similarly, the dominant charge state is confirmed to be 14+ (neodymiumlike) at 340~eV, and 12+ (samariumlike) at 280~eV.

Figure~\ref{fig:EUV_W}(a) shows the EUV spectra in the 22.5 -- 27~nm range observed at the same time with the TOF measurement shown in Fig.~\ref{fig:TOF_W}(a).
\begin{figure*}
\includegraphics[width=0.7\textwidth]{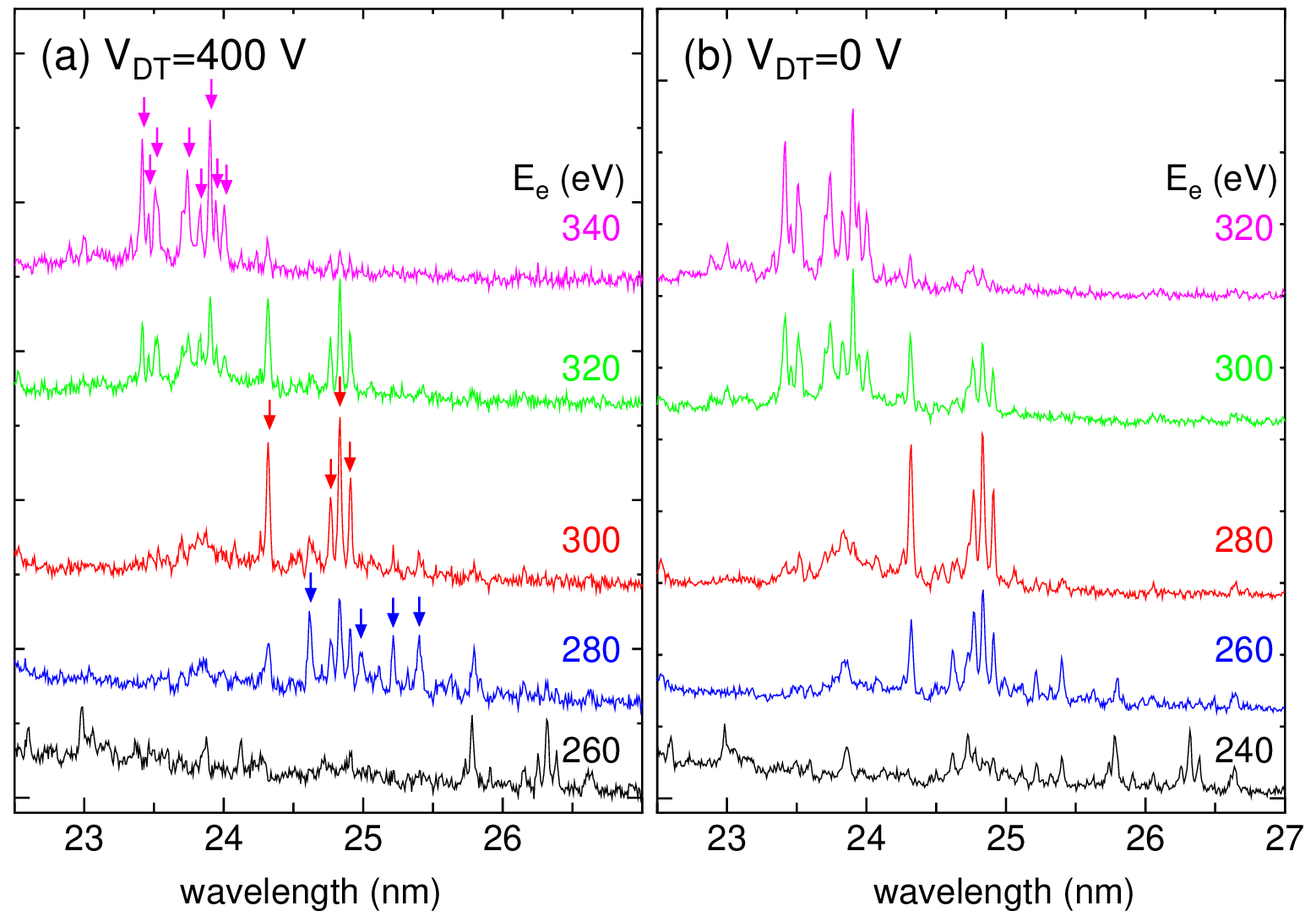}
\caption{\label{fig:EUV_W}
EUV spectra of W ions obtained with (a) the ion extraction mode where the potential of the ion trap ($V_{\rm DT}$) was biased to +400 V, and (b) the spectroscopy mode where $V_{\rm DT}$ was fixed at 0 V.
For example, for an electron energy of 260~eV, the cathode potential was +140 V for the ion extraction mode whereas it was -260 V for the spectroscopy mode.
The ion extraction mode spectra were obtained at the same time with observing TOF of extracted ions shown in Fig.~\ref{fig:TOF_W}(a).
}
\end{figure*}
\if0
\begin{figure}[t]
\includegraphics[width=0.49\textwidth]{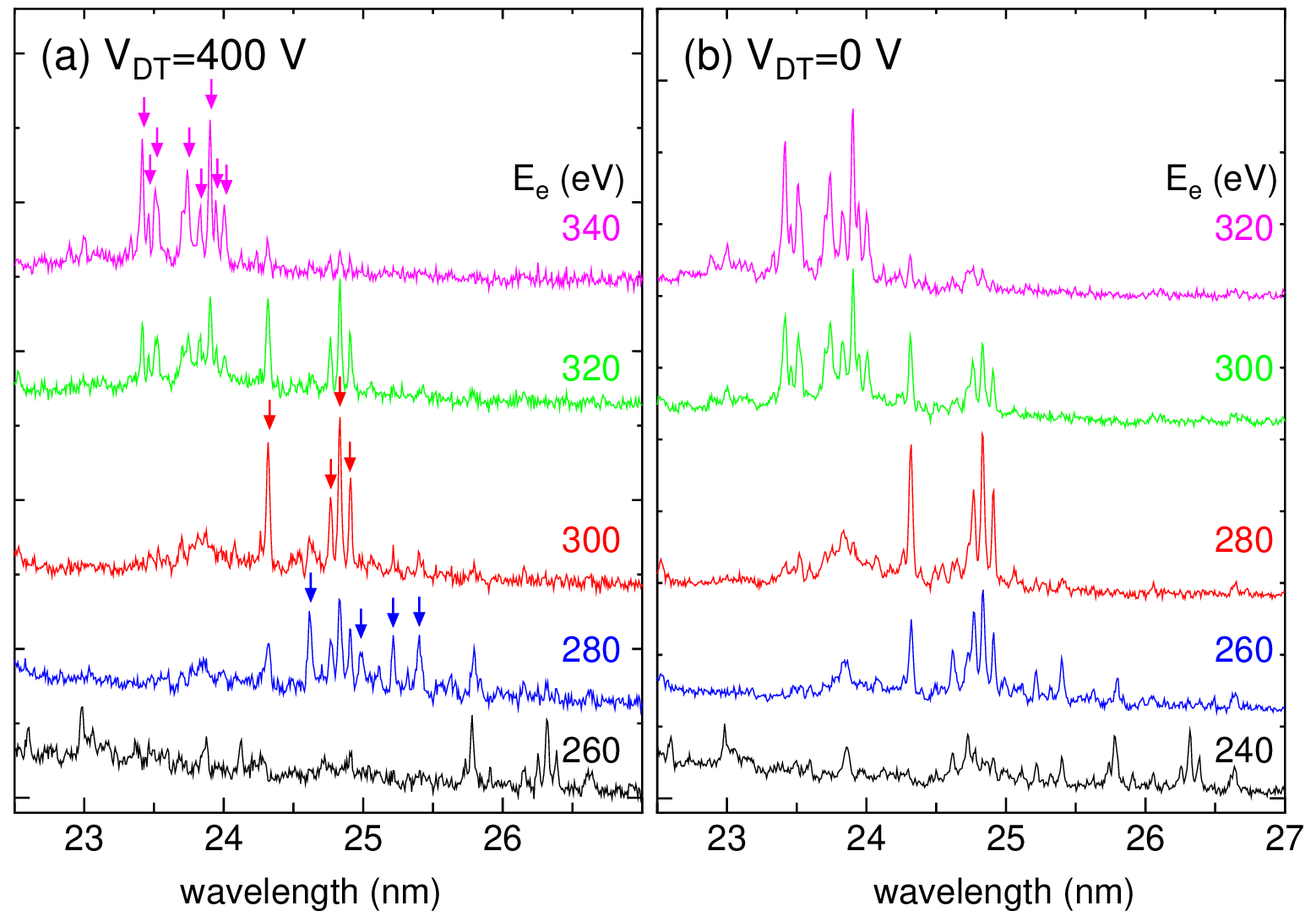}
\caption{\label{fig:EUV_W}
EUV spectra of W ions obtained with (a) the ion extraction mode where the potential of the ion trap ($V_{\rm DT}$) was biased to +400 V, and (b) the spectroscopy mode where $V_{\rm DT}$ was fixed at 0 V.
For example, for an electron energy of 260~eV, the cathode potential was +140 V for the ion extraction mode whereas it was -260 V for the spectroscopy mode.
The ion extraction mode spectra were obtained at the same time with observing TOF of extracted ions shown in Fig.~\ref{fig:TOF_W}(a).
}
\end{figure}
\fi
Similarly to Fig.~\ref{fig:TOF_W}, the electron energy represents the potential difference between the cathode and DT2.
Considering the TOF result, the lines that appeared at 280~eV, became dominant at 300~eV and almost disappeared at 340~eV are considered to be the lines from promethiumlike tungsten.
The lines showing such dependence, i.e. the lines that should be assigned to promethiumlike tungsten are indicated by the arrows in the spectrum at 300~eV.
Similarly, the lines that should be assigned to samariumlike (12+) and neodymiumlike (14+) tungsten are indicated by the arrows in the spectrum at 280~eV and 340~eV, respectively.

The spectra in Fig.~\ref{fig:EUV_W}(a) were observed with a drift tube potential of +400 V because the trapped ions should be extracted periodically for the TOF measurements.
On the other hand, Fig.~\ref{fig:EUV_W}(b) shows the spectra observed with the spectroscopy mode, i.e. with a grounded drift tube.
As confirmed from the comparison between (a) and (b), the electron energy dependence for each line differs from one another; the appearance energy is higher for the ion extraction mode by 20 -- 30~eV.
The reproducibility of this behavior was quite well, i.e. the higher appearance energy in the ion extraction mode was always confirmed regardless of the ion species.
This is probably due to the potential disturbance by the stainless tube installed inside the drift tube for gas injection.
The stainless tube is at the ground potential, which should make the potential in the trap lower when the drift tube is biased to a positive potential.
Thus the actual electron beam energy is considered to be lower than the electron energy values indicated in Fig.~\ref{fig:EUV_W}(a), which are just estimated from the difference in the output voltage of the power supplies used for biasing the cathode and the drift tube.
Although the electron energy values indicated in the spectroscopy mode should also be different from the actual energy due to the space charge potential of the electron beam and so on, they should be much closer to the actual energy.
Thus, judging from Fig.~\ref{fig:EUV_W}(b), the appearance energy for the lines assigned to promethiumlike W$^{13+}$ is around 260~eV and that assigned to neodymiumlike W$^{14+}$ is around 280~eV.
These appearance energies seem to be inconsistent with the ionization energy of tungsten ions, which is 258~eV for ionizing W$^{12+}$ to W$^{13+}$ and 291~eV for ionizing W$^{13+}$ to W$^{14+}$ according to Ref.~\cite{Kramida1}.
We consider this inconsistency arose from one of (or combination of) the following three reasons.
First, the ionization energy is not well known, so that it may be lower than the tabulated values.
Although an uncertainty of 1.2~eV is given in Ref.~\cite{Kramida1}, it may be difficult to calculate the ionization energy with such an accuracy especially for the ion having excited levels nearly degenerated with the ground state just as the present case.
Second, both samariumlike W$^{12+}$ and promethiumlike W$^{13+}$ have metastable states whose population can be comparable with or even much higher than that of the ground state.
According to the present calculation, the energy of such metastable states is 5 to 20~eV for W$^{12+}$ and 10 to 15~eV for W$^{13+}$ (see Fig.~\ref{fig:levels_pm}), which can make the appearance energy lower than the ionization energy estimated for the ground state.
Third, the electron energy in the ion trap could have a higher energy component due to the trap potential (30 V) applied to the both ends (DT1 and 3) of the drift tube.
Since the electrostatic trap potential in the present EBIT is not an ideal squared well potential, but more like a harmonic potential, the electron energy depends on the axial position in the trap.
We consider the charge identification based on the TOF measurement is much more reliable than the identification based on the appearance energy.
%According to the present result, promethiumlike tungsten has four prominent lines in this wavelength range, which are dominant in the spectrum at 300~eV.
%The spectrum is quite different from the previously observed spectra~\cite{Hutton2,Wu1} with the Berlin EBIT.
%At present we have no idea for this inconsistency.

\begin{figure}[t]
\includegraphics[width=0.4\textwidth]{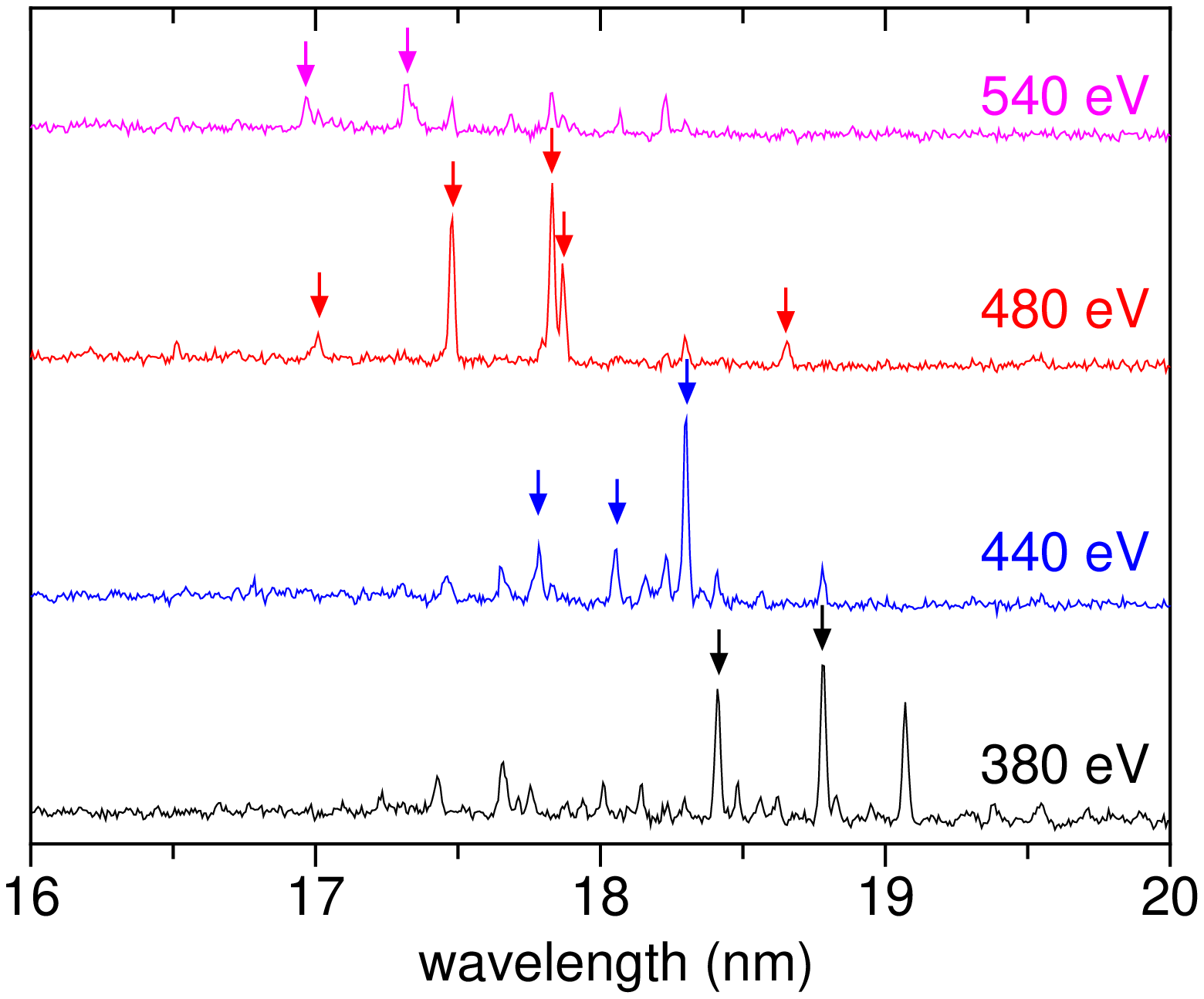}
\caption{\label{fig:EUV_Au}
EUV spectra of gold ions.
Note that the spectra were obtained at the ion extraction mode, so that the electron energy values shown in the figure should be higher than the actual ones (see text for the detail).
}
\end{figure}

Figure~\ref{fig:EUV_Au} shows the EUV spectra of highly charged gold ions for electron energies of 380 -- 540~eV.
Since these spectra were also obtained with the ion extraction mode with a drift tube potential of +400 V, the electron energy values indicated in the figure are considered to be higher than the actual ones.
In fact, observation with the spectroscopy mode were also made, and it was confirmed that the energy dependence shifted to lower energy by 20 -- 30~eV similarly to the tungsten observations.
The arrows in the spectra at 380, 440, 480, and 540~eV indicate the emission lines from europiumlike (16+), samariumlike (17+), promethiumlike (18+), and neodymiumlike (19+) gold ions, respectively, assigned based on the TOF analysis done at the same time with the spectroscopic observation.

Recently, Bekker {\it et al.}~\cite{Bekker1} observed the EUV spectra of promethiumlike ions with $Z=75-78$, which fill a gap between $Z=74$ and 79 in the present observation.
Their spectra obtained with the Heidelberg EBIT show a reasonable consistency with the present spectra of promethiumlike tungsten (spectrum at 300~eV in Fig.~\ref{fig:EUV_W}(a)) and promethiumlike gold (spectrum at 480~eV in Fig.~\ref{fig:EUV_Au}), i.e. the wavelength of the prominent transitions for $Z=74$ and 79 is in accordance with the $Z$-dependence observed with the Heidelberg EBIT for $Z=75-78$.
This would support the reliability of their and the present identification mutually.

\begin{figure}[t]
\includegraphics[width=0.4\textwidth]{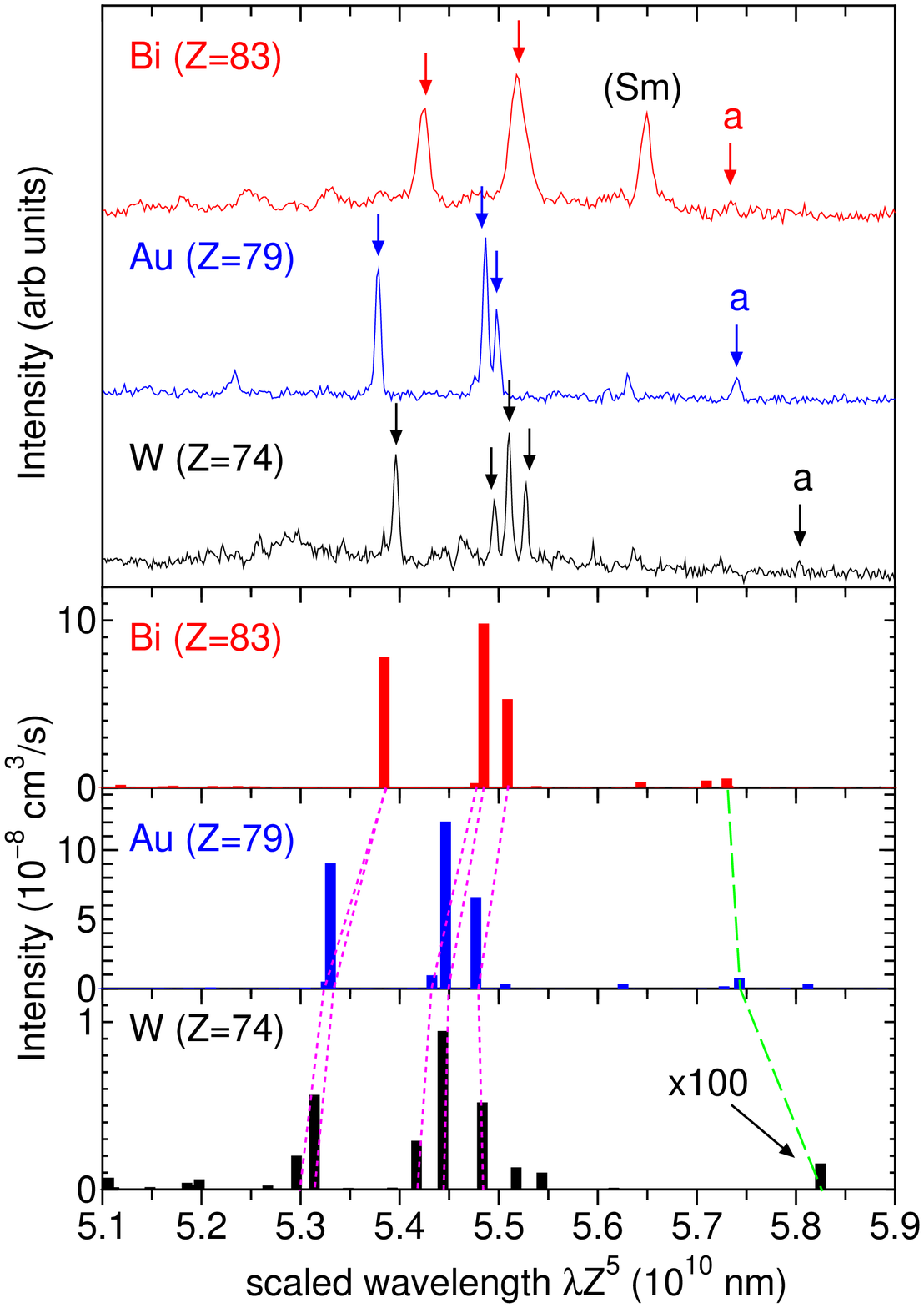}
\caption{\label{fig:pm-like}
Experimental and CR-model spectra for $5s$ -- $5p_{3/2}$ transitions of promethiumlike tungsten, gold, and bismuth.
The horizontal axis is wavelength ($\lambda$) scaled by $Z^5$ ($\lambda \cdot Z^5$ [$10^{10}$~nm]).
}
\end{figure}

Figure~\ref{fig:pm-like} shows the EUV spectra of promethiumlike tungsten, gold, and bismuth~\cite{Kobayashi1} compared with the present CR model spectra obtained for an electron density of $10^{10}$ cm$^{-3}$.
The prominent lines in the model spectra, which are connected by dotted lines, correspond to the $4f^{13}5s^2$ -- $4f^{13}5s5p_{3/2}$ transitions.
Although the ground state configuration is $4f^{14}5s$ for gold and bismuth, the population of the metastable $4f^{13}5s^2$ states dominates over that of the ground state (see Fig.~\ref{fig:levels_pm} and Sec.~\ref{sec:theory}), and as a result, the $4f^{13}5s^2$ -- $4f^{13}5s5p$ transitions are much more prominent compared with the resonant $4f^{14}5s$ -- $4f^{14}5p$ transition as already shown in our previous study~\cite{Kobayashi1}.
%According to the CR model calculation, the relative population of the metastable [$4f_{7/2}^{-1}5s5p$]$_{J=7}$ state with respect to that of the ground state is 11 for gold and 15 for bismuth at an electron density of $10^{10}$ cm$^{-3}$, which is the typical density of CoBIT~\cite{Kobayashi1,Nakamura21}.
On the other hand, $4f^{13}5s^2$ is the ground state for tungsten (see Fig.~\ref{fig:levels_pm}); thus the $4f^{13}5s^2$ -- $4f^{13}5s5p$ transitions correspond to the resonance lines for promethiumlike tungsten.
However, $4f^{12}5s^2 5p$ states can have a population comparable to that of the ground state as already discussed in Sec.~\ref{sec:theory}.
%For example, according to the CR model calculation, the population of the [$4f_{5/2}^{6}4f_{7/2}^{6}5s^2 5p_{1/2}$]$_{J=7/2-13/2}$ states with a level energy of 10 -- 12~eV is a few times larger than that of the $4f^{13}5s^2$ ground state.
As a result, the intensity of the $4f^{13}5s^2$ -- $4f^{13}5s5p$ transitions for tungsten is about an order of magnitude smaller than that for gold and bismuth as shown in the CR model spectra in Fig.~\ref{fig:pm-like} even though they are the resonance lines for tungsten.
As seen in Fig.~\ref{fig:pm-like}, overall agreement is confirmed between the experimental and model spectra although slight shift in wavelength exists.
%The consistency between the experiment and theory would support the reliability of the present identification of the charge state.
From the comparison, the lines indicated by the arrows are assigned as $4f^{13}5s^2$ -- $4f^{13}5s5p$ transitions, and their wavelengths are listed in Table~\ref{tab:euv}.
\begin{table*}
\caption{\label{tab:euv}Experimental and theoretical wavelengths ($\lambda_{\rm ex}$ and $\lambda_{\rm th}$ in nm) and theoretical transition probabilities ($A$ in $10^{11}$~s$^{-1}$) for [$4f^{13}5s^2$]$_{J_f}$ -- [$4f^{13}5s5p_{3/2}$]$_{J_i}$ transitions in promethiumlike ions.
Note that the $4f^{13}$ core is not changed during the transition; thus the angular momentum of the $4f$ vacancy of the upper level is the same as $J_f$.
%The notation $a(b)$ for the transition probability denotes $a \times 10^b$.
}
\begin{ruledtabular}
\begin{tabular}{cccccccccccc}
&&\multicolumn{4}{c}{W ($Z=74$)}&\multicolumn{3}{c}{Au ($Z=79$)}&\multicolumn{3}{c}{Bi ($Z=83$)}\\
\cline{3-6} \cline{7-9} \cline{10-12}\\[-3mm]
$J_i$&$J_f$&$\lambda_{\rm th}^{\rm a}$&$\lambda_{\rm th}^{\rm b}$&$\lambda_{\rm ex}$&$A$&$\lambda_{\rm th}^{\rm a}$&$\lambda_{\rm ex}$&$A$&$\lambda_{\rm th}^{\rm a}$&$\lambda_{\rm ex}^{\rm c}$&$A$\\
\hline
5/2&5/2&23.87&24.00&--&0.63&17.31&--&1.3&13.67&--&2.1\\
7/2&7/2&23.95&24.06&24.32&0.54&17.32&17.48&1.3&13.67&13.77&2.1\\
7/2&5/2&24.41&24.57&24.77&0.51&17.66&--&1.2&13.90&--&2.0\\
9/2&7/2&24.53&24.64&24.83&0.61&17.70&17.83&1.2&13.92&\multirow{2}{*}{14.0*}&2.0\\
5/2&7/2&24.71&24.70&24.91&0.54&17.80&17.87&1.1&13.98&&1.8\\
\end{tabular}
\end{ruledtabular}
\leftline{*Blend. $^{\rm a}$Present results with HULLAC. $^{\rm b}$Hartree-Fock-relativistic method (COWAN code)~\cite{Safronova2}. $^{\rm c}$Ref.~\cite{Kobayashi1}}.
\end{table*}

In the model spectra shown in Fig.~\ref{fig:pm-like}, the weak peaks connected by a dashed line correspond to the $4f^{14}5s$ -- $4f^{14}5p_{3/2}$ transition (note that the intensity is multiplied by 100 only for tungsten).
From the comparison with the model, the weak peaks labeled ``a" in the experimental spectra seem to correspond to it.
Thus we tentatively identify these peaks as the $4f^{14}5s$ -- $4f^{14}5p_{3/2}$ transition for bismuth and gold.
For bismuth, the peak intensity is too weak to determine its attribution; thus it can be the line from other charge states of bismuth or even from an impurity ion.
For gold, it is definitely identified as a line from promethiumlike based on the energy dependence (Fig.~\ref{fig:EUV_Au}) and the TOF analysis; thus the identification is probably correct.
On the other hand for tungsten, the weak peak was also observed at lower electron energies; thus it is considered to be a line from a tungsten ion with a lower charge.
According to the calculation, the $4f^{14}5s$ state is an excited state for tungsten, whose lifetime is short as $10^{-4}$ s.
Thus its population is very small (in the order of $10^{-4}$ with respect to the ground state) so that there would be no chance to observe the $4f^{14}5s$ -- $4f^{14}5p$ transition for promethiumlike tungsten at least with an EBIT.
The wavelengths of the tentatively identified $4f^{14}5s$ -- $4f^{14}5p_{3/2}$ transition are listed in Table~\ref{tab:res} together with present and existing calculations.
\begin{table*}
\caption{\label{tab:res}
Wavelengths of the $5s$ -- $5p$ transitions in promethiumlike ions. Transition probability $A$ (in $10^{11}$ s$^{-1}$) is given only for the present result with HULLAC.
}
\begin{ruledtabular}
\begin{tabular}{cccccccc}
&$Z$&Present (Th)&HFR$^{\rm a}$&MR-MP$^{\rm b}$&DF$^{\rm c}$&Present (Ex)&$A$\\
\hline
%$5s$--$5p_{1/2}$&&&&&&&\\
$5s$--$5p_{1/2}$&74&36.949&36.795&36.914&37.079&&0.13\\
&79&28.980&29.115&29.149&29.033&&0.20\\
&83&24.508&24.719&&&&0.27\\
%$5s$--$5p_{3/2}$&&&&&&&\\
$5s$--$5p_{3/2}$&74&26.249&26.083&25.962&26.242&&0.38\\
&79&18.663&18.566&18.614&18.643&18.66*&0.77\\
&83&14.547&14.424&&&14.56*&1.3\\
\end{tabular}
\end{ruledtabular}
\leftline{$^{\rm a}$Hartree-Fock-relativistic method (COWAN code)~\cite{Safronova2}. $^{\rm b}$Multireference M$\phi$ller-Plesset (MR-MP) theory~\cite{Vilkas1}.}
\leftline{$^{\rm c}$Single configuration Dirac-Fock calculation~\cite{Theodosiou1}. *Tentative identification.}
\end{table*}

\begin{figure}[t]
\includegraphics[width=0.45\textwidth]{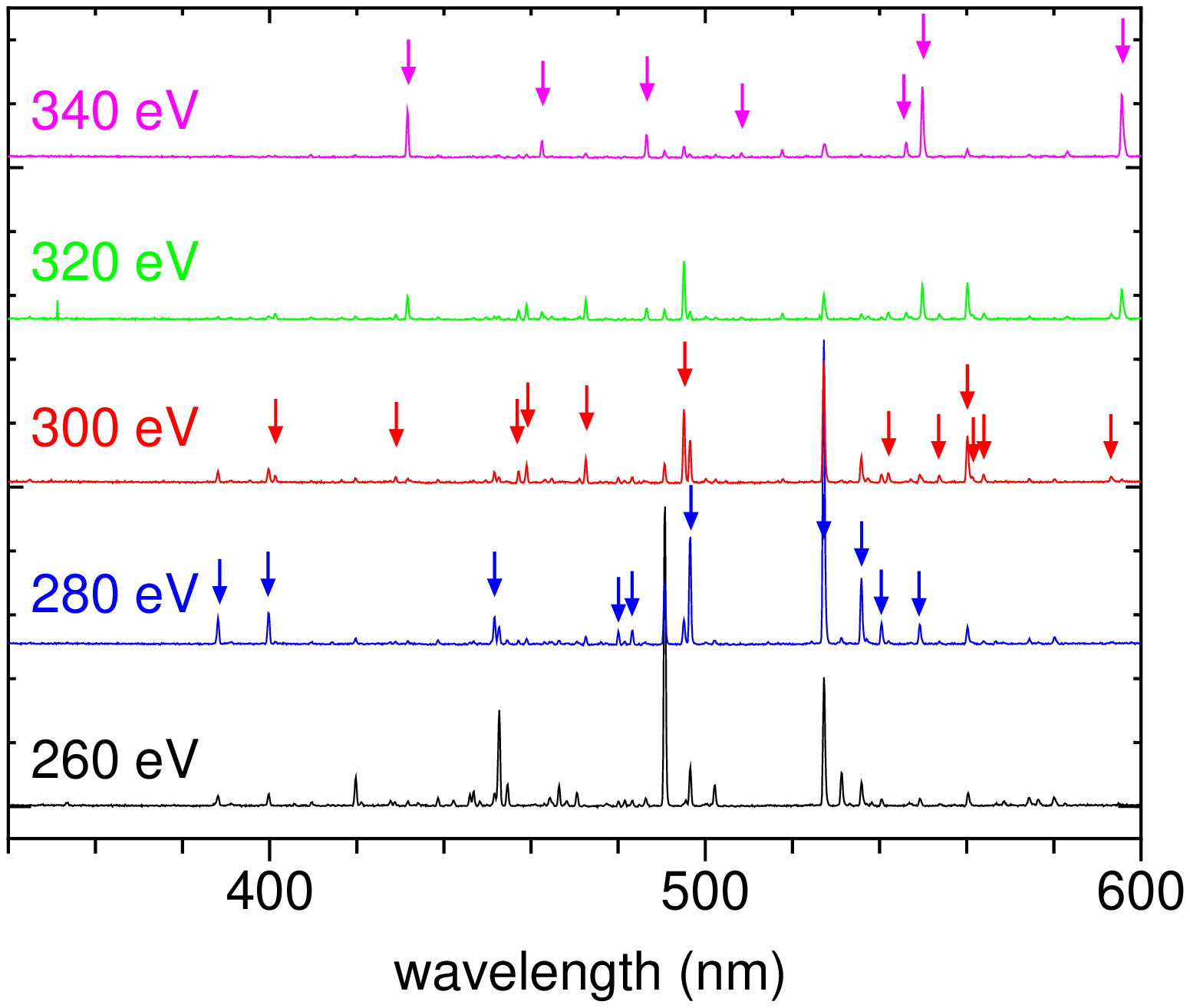}
\caption{\label{fig:vis_W_wide}
Visible spectra of tungsten ions for the 340 to 600~nm range obtained while observing TOF of extracted ions shown in Fig.~\ref{fig:TOF_W}(a) and EUV spectra shown in Fig.~\ref{fig:EUV_W}(a).
The arrows in the spectra at 280, 300, and 340~eV indicate the emission lines of samariumlike, promethiumlike, and neodymiumlike ions, respectively, identified through the TOF of the extracted ions.
}
\end{figure}

Figure~\ref{fig:vis_W_wide} shows visible spectra of tungsten ions observed at the same time with the TOF shown in Fig.~\ref{fig:TOF_W}(a) and the EUV spectra shown in Fig.~\ref{fig:EUV_W}(a).
A grating with a groove number of 300~mm$^{-1}$ was used to observe a wide wavelength range.
The spectra were obtained in the ion extraction mode (i.e. with a biased drift tube); thus the electron energy values indicated in the figure are the identical with those in Fig.~\ref{fig:EUV_W}(a), which should be higher than the actual ones.
Based on the TOF measurement, the lines indicated by the arrows in the spectra at 280, 300, and 340~eV have been identified as the emission lines of samariumlike, promethiumlike, and neodymiumlike tungsten, respectively.
Most of the lines observed in Fig.~\ref{fig:vis_W_wide} were recently observed with a compact EBIT in Shanghai~\cite{Zhao1}.
Although the energy dependence is quite similar between their and our present observations, their identification of the charge state differs from our present identification by unity, i.e. the lines that we assigned to neodymiumlike W$^{14+}$ have been assigned to promethiumlike W$^{13+}$ in their study. 
They identified the charge state based on the appearance energy comparing it with the theoretical ionization energies tabulated in Ref.~\cite{Kramida1}.
Probably, their identification was made based on the fact that the corresponding lines appeared at an electron energy below the ionization energy of W$^{13+}$.
However, we consider that our present identification based on the TOF analysis is reliable and definite.
Recently, the visible spectrum of neodymiumlike W$^{14+}$ was also observed with the Heidelberg EBIT~\cite{Windberger1}.
The spectrum and the identification in their study is consistent with those in the present study.

\begin{figure}[t]
\includegraphics[width=0.4\textwidth]{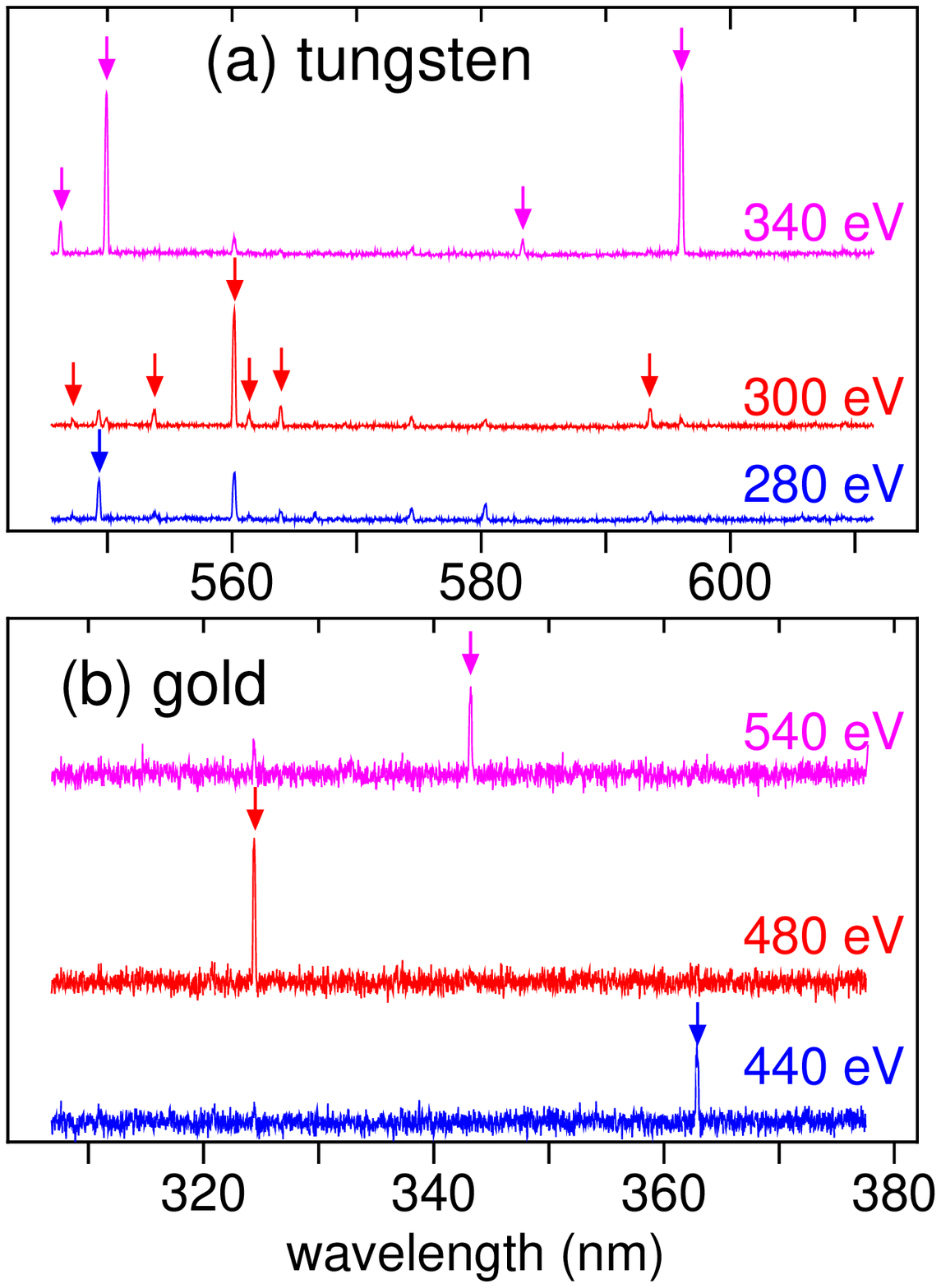}
\caption{\label{fig:vis_1200}
High resolution visible spectra of (a) tungsten and (b) gold, obtained while observing TOF of extracted ions.
The arrows in the spectra at 280, 300, and 340~eV indicate the emission lines of samariumlike, promethiumlike, and neodymiumlike tungsten, respectively, and those at 440, 480, and 540~eV indicate lines of samariumlike, promethiumlike, and neodymiumlike gold, respectively.
}
\end{figure}

High resolution visible spectra were also observed with a 1200~mm$^{-1}$ grating to determine the wavelength of the observed lines listed in Table~\ref{tab:vis}.
\begin{table}
\caption{\label{tab:vis}Wavelength (in air) of the observed visible transitions in tungsten ions in the wavelength range of 380 to 600~nm.
The values without superscript are obtained in this study, whereas the values with superscript are taken from our previous study~\cite{Komatsu2}, and reconfirmed in the present study.
}
\begin{ruledtabular}
\begin{tabular}{ll}
ion&wavelength (nm)\\
\hline
W$^{12+}$&388.19$^{\rm a*}$, 399.81$^{\rm a*}$, 451.68$^{\rm a}$, 480.09, 483.26, 496.55,\\
&527.27, 535.90, 540.53, 549.33\\
W$^{13+}$&401.38$^{\rm a*}$, 429.03, 457.26$^{\rm a}$, 459.08$^{\rm a}$, 472.68$^{\rm a}$, 495.16,\\
&537.49, 542.11, 547.22, 553.81, 560.25, 561.46,\\
&563.99, 593.28\\
W$^{14+}$&431.75, 462.59$^{\rm a}$, 486.57, 506.40, 508.39, 517.74,\\
&527.70, 546.22, 549.93, 583.23, 595.70\\
\end{tabular}
\end{ruledtabular}
\leftline{$^{\rm a}$Data taken from our previous study~\cite{Komatsu2}.}
\leftline{$^{\rm *}$Assignment modified from Ref.~\cite{Komatsu2}.}\\
\end{table}
It is noted that the assignment in our previous study is modified for a few lines.
An example for the high resolution spectra of tungsten ions is shown in Fig.~\ref{fig:vis_1200}(a) for the 540 -- 610~nm range, where the M1 transition between the ground state fine structure levels is expected to be observed.
Among the six lines observed in this wavelength range, the most prominent line near 560~nm is tentatively identified to be the M1 transition in the ground state.
In Table~\ref{tab:vis_comp}, the experimental wavelength is shown with the present theoretical results of the wavelengths and the M1 transition rates ($A$ coefficients) calculated using GRASP2K.
%For the present calculations, single reference CSF was included in the expansion of Eq.~(\ref{eq:MCDHF}) for each level, i.e. $4f^{14}5s$, $4f^{14}5p$ $J=1/2$, 3/2, respectively.
\begin{table}
\caption{\label{tab:vis_comp}Wavelength $\lambda$ (nm in air) and transition probability $A$ (s$^{-1}$) for the [$4f^{13}5s^2$]$_J$ ($J=7/2$ -- 5/2) transition in promethiumlike tungsten and gold.
}
\begin{ruledtabular}
\begin{tabular}{lllllll}
ion&$\lambda_{\rm exp}$&$\lambda_{\rm th}$&&&&$A$\\
\hline
W$^{13+}$&560.25&567.8$^{\rm a}$&538$^{\rm b}$&568$^{\rm c}$&552$^{\rm d}$&83.9$^{\rm a}$\\
Au$^{18+}$&324.39&327.4$^{\rm a}$&&&&437$^{\rm a}$\\
\end{tabular}
\end{ruledtabular}
\leftline{$^{\rm a}$Present result with GRASP2K.}
\leftline{$^{\rm b}$Hartree-Fock relativistic calculation with COWAN\cite{Safronova2}}
\leftline{$^{\rm c}$Relativistic many-body perturbation theory with FAC\cite{Zhao1}.}
\leftline{$^{\rm d}$Relativistic configuration interaction with FAC\cite{Zhao1}.}\\
\end{table}
The present calculations of the fine structure splitting for W$^{13+}$ is summarized in Table~\ref{tab:grasp} showing trend of the calculated values along the number of CSFs in the expansion of Eq.~(\ref{eq:MCDHF}).
\begin{table}
\caption{\label{tab:grasp}Fine structure splitting (cm$^{-1}$) of the $4f^{13}5s^2$ ground state in promethiumlike tungsten calculated with GRASP2K.
The second and third column indicate the number of CSFs in the expansion for each active set.
}
\begin{ruledtabular}
\begin{tabular}{lrrl}
active set&$J=5/2$&7/2&splitting\\
\hline
reference&1&1&17384.47\\
SD5&145&155&17397.14\\
SD5+4f&3470&3927&17484.02\\
SD5+4df&15363&17270&17574.50\\
SD5+4pdf&28144&31421&17607.29\\
\end{tabular}
\end{ruledtabular}
\end{table}
Each set of the CSFs in the expansion is generated by single and double excitations from orbitals in the reference configuration [Kr]$4d^{10}4f^{13}5s^2$ using the active set method~\cite{Olsen1,Sturesson1}.
The active sets of orbitals are indicated by SD5 for $5l$ ($l=0$ -- 4), SD5+$4f$ for $5l$ and $4f$, SD5+$4df$ for $5l$, $4d$, and $4f$, and SD5+$4pdf$ for $5l$, $4p$, $4d$, and $4f$, respectively.
The calculated splitting becomes larger and getting closer to the experimental value as the number of CSFs increases.
The corresponding wavelength obtained with the largest active set, i.e. SD5+$4pdf$, is in reasonable agreement with the measured values (see table~\ref{tab:vis_comp}).

Since the ground state of promethium gold is $4f^{14}5s$, there should not be M1 transitions in the ground state.
However, if the population of the metastable $4f^{13}5s^2$ state is large enough, there would be a chance to observe the M1 transition between its fine structure levels $J=7/2$ and 5/2 in the visible range.
%According to our calculation, the wavelength of the M1 transition is 332~nm.
Figure~\ref{fig:vis_1200}(b) shows the visible spectra of gold ions obtained for the wavelength region where the M1 transition is expected to be observed.
The prominent line at 324~nm observed at an electron energy of 480~eV is assigned to promethiumlike gold based on the TOF analysis performed at the same time with the spectroscopic observation.
Thus it is probably the M1 transition between the fine structure levels in the metastable state.
The experimental wavelength for gold is also listed in Table~\ref{tab:vis_comp} and compared with the present theoretical result, which shows reasonable agreement.
The lines (362.88~nm and 343.19~nm) observed at 440~eV and 540~eV are assigned to samariumlike and neodymiumlike gold, respectively, although the detailed transition is not identified in this study.

\section{\label{sec:conclusion}Conclusion}

We have observed extreme ultraviolet and visible spectra of highly charged tungsten and gold ions with a compact electron beam ion trap (EBIT) in Tokyo.
In order to identify the charge state of the ion which should be assigned to the observed lines, the time-of-flight (TOF) charge analysis was done at the same time with the spectroscopic observation.
Some previous ambiguous identifications in promethiumlike tungsten and gold have been clarified owing to the define identification by the TOF analysis.
Our present study has proved that an EBIT is a powerful tool to observe and identify previously-unreported lines of highly charged heavy ions, but also showed that careful consideration on the metastable state should be taken into account.
In particular, many electron systems near closed shell configuration, just as promethiumlike ions, population trapping to metastable states can affect the structure and the electron energy dependence of observed spectra considerably.

\section*{Acknowledgment}
%This work was supported by JSPS KAKENHI Grant Numbers 23246165 and 15H04235, and partly supported by the JSPS-NRF-NSFC A3 Foresight Program in the field of Plasma Physics (NSFC: No.11261140328).

This work was performed with the support and under the auspices of the NIFS Collaboration Research program (NIFS09KOAJ003) and JSPS KAKENHI Grant Number 23246165 and 15H04235, and partly supported by the JSPS-NRF-NSFC A3 Foresight Program in the field of Plasma Physics (NSFC: No.11261140328, NRF: 2012K2A2A6000443).

\bibliography{ref}

\end{document}